%% file: conference_101719.tex
\documentclass[conference]{IEEEtran}
\IEEEoverridecommandlockouts
\usepackage[utf8]{inputenc}
\usepackage{cite}
\usepackage{comment}
\usepackage{graphicx, floatrow}
\usepackage{graphicx}
\usepackage[caption=false,font=footnotesize]{subfig}
\usepackage{cite}
\usepackage{caption}
\usepackage{amsmath,amssymb,amsfonts}
\usepackage{algorithmic}
\usepackage{graphicx}
\usepackage{textcomp}
\usepackage{xcolor}
\usepackage{floatrow}
\usepackage{tikz}
\usepackage{makecell}
\def\BibTeX{{\rm B\kern-.05em{\sc i\kern-.025em b}\kern-.08em
    T\kern-.1667em\lower.7ex\hbox{E}\kern-.125emX}}
    
\begin{document}

\title{The Effects of Correctly Modeling Generator Step-Up Transformer Status in Geomagnetic Disturbance Studies\\
\thanks{
The authors would like to acknowledge the support of the U.S. Department of Energy, Office of Cybersecurity, Energy Security, and Emergency Response (DOE-CESER) and the Pacific Northwest National Laboratory (PNNL) for work conducted under Master Agreement 401273, Task Order 610876.\\
\\
Copyright \copyright 2022 IEEE. Personal use of this material is permitted. However, permission to use this material for any other purposes must be obtained from the IEEE by sending a request to pubspermissions@ieee.org. Presented at the 2022 North American Power Symposium (NAPS), Salt Lake City, UT, October 2022.}
}

\author{\IEEEauthorblockN{Jessica L. Wert, Pooria Dehghanian, Jonathan Snodgrass, Thomas J. Overbye}
\IEEEauthorblockA{\textit{Department of Electrical and Computer Engineering} \\
\textit{Texas A\&M University}\\
College Station, TX, USA \\
\{jwert, pooria.dehghanian,  snodgrass, overbye\}@tamu.edu}
}

\maketitle

\begin{abstract}

In order to correctly model the impacts of geomagnetically induced current (GIC) flows, the generator step-up (GSU) transformer status must be properly modeled. In power flow studies, generators are typically removed from service without disconnecting their GSU transformers since the GSU transformer’s status has little to no impact on the power flow result. In reality, removing a generator from service involves also removing its GSU transformer from service. This difference presents a discrepancy between simulated behavior and system observations during geomagnetic disturbance (GMD) events. This paper presents GMD case studies on 2000-bus and 24,000-bus systems in which reactive power losses and geomagnetically induced currents are compared across scenarios in which the GSU transformers for the disconnected generators are either in-service or out-of-service. The results demonstrate a 3.2\textemdash15.5\% error in  reactive power losses when the GSU status is modeled incorrectly. Discrepancies of up to 95 A per phase for branch GIC flows and 450 A for transformer neutral GIC flows are also observed and visualized.  

\end{abstract}
\vspace{3mm}



\section{Introduction}
\input{Sections/I_Introduction}

\section{Background}
\label{sec:Background}

\input{Sections/II_Background}

\section{Case Study}
\label{sec:CaseStudy}

\input{Sections/III_CaseStudy}

\section{Results}
\label{sec:Results}

\input{Sections/IV_ResultsAnalysis}


\section{Summary}
\label{sec:Summary}

\input{Sections/V_SummaryFutureWork}

\bibliographystyle{IEEEtran}
\bibliography{bibi.bib}

\end{document}

%% file: Sections/I_Introduction.tex
\IEEEPARstart{G}{eomagnetic} disturbances (GMDs) are caused by large bursts of intense magnetic particles released by strong solar storms. A GMD is characterized by variations in the earth's magnetic field at its surface which produce low-frequency quasi-dc currents running along transmission lines and through transformer windings \cite{GIC_Model3,Load-Flow-GIC}. The flow of these geomagnetically-induced currents (GICs) through the system is related to available paths to ground, such as grounded transformers \cite{Overbye1}. To correctly simulate GICs, the status of generator step-up (GSU) transformers should be properly represented in the power system models. Understanding GICs is useful because they can subject transformers to half-cycle saturation, which results in the creation of harmonics, increased reactive power losses, and heating of transformer windings and other internal components \cite{GIC_Trans,TPEC}. Power grid equipment such as high-voltage power transformers may suffer serious effects from GMDs \cite{nerc_gmd,GIC_model1,GIC_Model2}. 


This paper has three main premises with most of the focus placed on the first and third. First, in order to correctly model the impact of GICs on an electric grid using power flow or transient stability analysis it is important to correctly model the statuses of the GSU transformers for the out-of-service generators. Second, since the GSU status of out-of-service generators does not impact the steady state power flow results, it is quite common to model most to all of the GSU transformers as in-service. Third, while there are typically not flags to indicate which transformers are GSU transformers, they can be identified using simple and efficient algorithms.

The rest of the paper is structured as follows: Section \ref{sec:Background} introduces an overview of GSU transformers, their modeling during steady state power flow calculations, and their algorithmic identification. The case studies, grid statistics, and developed scenarios are presented in Section \ref{sec:CaseStudy}. The simulation analysis and numerical results are presented in Section \ref{sec:Results}. Lastly, a summary is provided in Section \ref{sec:Summary}.

%% file: Sections/II_Background.tex
\subsection{Generator Step-Up Transformers in Power Systems}
\label{sec:GSUs-Power-Systems}
\begin{figure}[t!]
    \centering
    \includegraphics[width =0.75\linewidth]{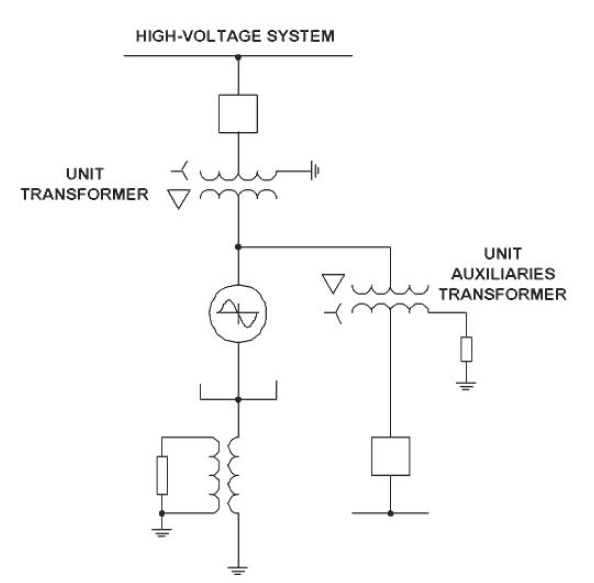}
    \caption{\color{black} IEEE unit generator-transformer configuration,\\ reproduced Figure 3-7 from \cite{IEEEC37}.}
    \label{fig:IEEEGSU transformerConfig}
\end{figure}

In a typical GSU installation, the delta-configured low voltage windings are connected to the generator terminals and the Y-configured high voltage windings are connected to the transmission line or substation \cite{banovic2014classification}. The IEEE guide for generator protection indicates that a circuit breaker must be placed at the high-side of the transformer as a minimum requirement \cite{IEEEC37}, shown in Figure \ref{fig:IEEEGSU transformerConfig}. However, per \cite{IEEE-GB-37-013-2021}, it is more common for larger generators to have a low-side breaker placed between the generator and the tapped connection to the unit auxiliary transformer. This breaker is termed the generator breaker and allows the GSU transformer to be rapidly isolated in the event of an internal GSU fault to prevent damage to the generator. 

For generator configurations with both high-side and low-side GSU breakers, it is still common to open both breakers when a generator is shut down or tripped \cite{IEEEC37}. However, in a typical bus-branch power flow model, the GSU transformers are rarely taken out of service even though the generator is off-line in the model, as explained below in Section \ref{sec:GSU-PowerFlow-Modeling}.

\subsection{GSU Modeling in Steady State Power Flow Calculations}
\label{sec:GSU-PowerFlow-Modeling}

To address the second premise that the GSU status doesn't affect the steady state power flow results, the generator and GSU statuses were examined for the real-world 62,600-bus and 8300-generator power flow model of the North American Eastern Interconnect (EI). This case is used in \cite{Overbye1} and represents a 2010 planning case for the then-anticipated peak 2012 conditions. The total generation is 893 GW with 159 GW modeled as out-of-service. Of the out-of-service generators, the generator buses are energized for all but 15 GW, indicating that the majority of the the generators' GSUs are in-service. Similar trends are seen in the 2019 series models used in \cite{Overbye2} with models for both the EI and the WECC considered with high- and low-load scenarios for both. The EI models have 87,000 buses and 9400 generators, whereas the WECC ones have 23,000 buses and 4300 generators. In the EI heavy-load case there is 917 GW of generation with 168 GW out-of-service yet with only 14.2 GW (8.5\%) at out-of-service buses. In the EI lower load case there is 932 GW of generation with 508 GW out-of-service yet with only 13.0 GW (2.6\%) at out-of-service buses. For the WECC high load case the values are 285 GW, 61.4 GW, and 18.1 GW (29.6\%) respectively. For the WECC lower load case the values are 282 GW, 151 GW, and 6.6 GW (4.8\%).

The point of showing these values is not to imply that they do not represent valid power flow cases. Since the GSU transformers for out-of-service generators have no flow and essentially no shunt reactive values, their statuses have essentially no impact on standard power flow solutions. However since GSUs are commonly grounded-wye on the high side and delta on the low side, GICs still flow through the transformer neutral connections and cause reactive power consumption as demonstrated in this paper; thus having the GSU statuses correct is important for GIC studies.

\subsection{GSU Transformer Identification Algorithm}

The paper’s final premise is that the identity of the out-of-service generator GSU transformers can usually be quickly determined algorithmically using the following proposed straightforward algorithm. The input to the algorithm is the electric grid topology used in the power flow (including the bus nominal voltage values) and an assumed minimum transmission level voltage (with 40 kV used as a default here). Then for each generator not directly connected to a transmission level bus the algorithm uses a using a breadth first search to determine the set of transformers connecting the generator to transmission level buses. Computationally the effort required per generator bus depends on its topology.  In the simplest situation, considering of a set of generators at the same bus directly connected to a transmission bus through a set of GSUs, this search is trivial. That is, just the single bus.  The worst case scenario is when there is no transmission level bus connection, such as when the grid is entirely at the distribution level, since the algorithm would need to search the entire grid. However, the situation can be managed by adding a maximum bus counter with the algorithm terminating if the counter is reached; here the counter is set to 20.

\begin{figure}[h!]
    \centering
    \includegraphics[width = 0.8\textwidth]{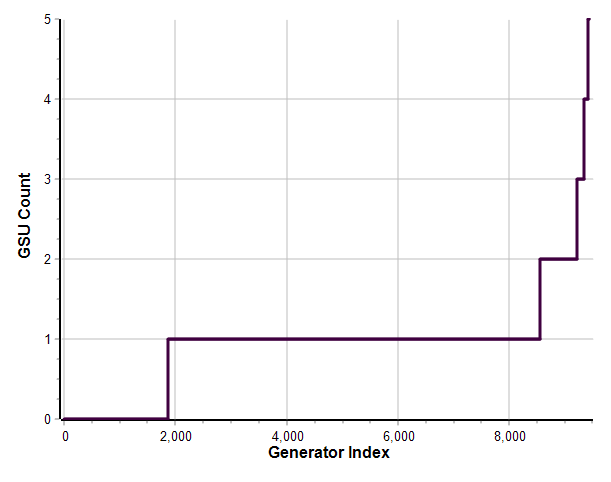}
    \caption{Number of algorithmically determined step-up transformers for an 87,000-bus EI model.}
    \label{fig:GSUs_87K}
\end{figure}

As an example, consider the previously-mentioned 87,000-bus, 9453-generator EI grid with Figure \ref{fig:GSUs_87K} showing the distribution of GSUs found. For the 1870 generators with no GSUs shown, 1298 are directly connected at voltages above the transmission threshold and not considered further (with many of these being small generators, reactive control devices, HVDC line equivalents, or future projects). Then 572 are in sub-transmission or distribution grids with no direct GSU path to the transmission grid. For the rest, the vast majority have one or two GSUs, making it straightforward for the person doing the study to determine whether they should be opened.


%% file: Sections/III_CaseStudy.tex
\subsection{Grid Models}

To demonstrate the first premise presented in this paper, scenarios are created from two different synthetic grids of varying sizes. The network models of each grid reflect real-world power systems \cite{SynthGrid1, SynthGrid2, SynthGrid3} and were obtained from \cite{O24k3}. Both of the grids modeled contain the necessary parameters for a GMD study, and each have GSU transformers that are explicitly modeled in the power flow.

The first case study grid is the ACTIVSg2000 synthetic grid with a geographic footprint located in Texas. It features a transmission network comprised of 500-kV, 230-kV, 161-kV, and 115-kV transmission lines. Of the 861 transformers within the case, 457 are GSU transformers.

\begin{figure}[h]
    \includegraphics[width=0.75\textwidth]{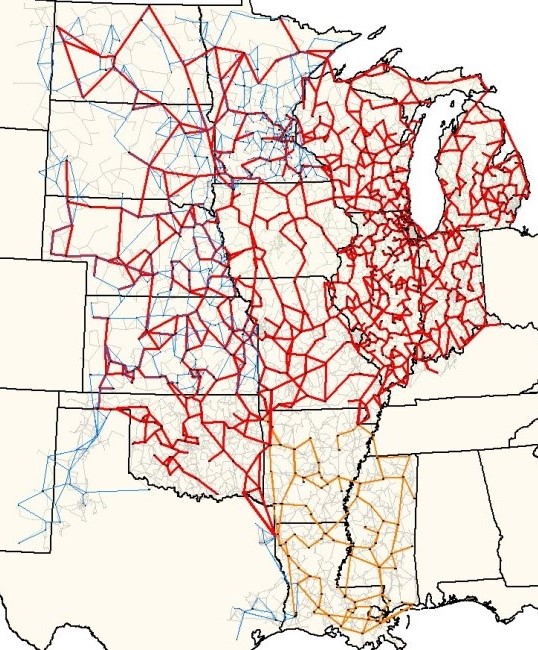}
    \caption{24,000-bus one-line diagram.}
    \label{fig:24KOneline}
\end{figure}

The second case study example is the 24,000-bus and 6300-generator grid from \cite{O24k1} that geographically covers a central region of the United States and is similar to the footprint of MISO and SPP combined. The one-line diagram for the grid is shown in Figure \ref{fig:24KOneline} with the nominal voltage of the transmission lines show using different colors with orange for 500 kV, red for 345 kV, blue for 230 kV, and black or gray for the lower voltages. Overall the grid has a total installed capacity of about 320 GW of generation with the size and location of the generators based on the 2019 EIA 860 data \cite{O24k2}. This grid is divided into 19 operating areas roughly corresponding to the US states with the number of buses per area ranging from 3670 down to 58. 

\subsection{Case Study Scenarios}
Simulations are performed on the 2000-bus case for low and high loading conditions. Here, the low load case represents a prototypical low load spring case with a load of 23 GW and the high load represents a peak summer load condition with a load of 66 GW. The load and unit commitment scenarios are borrowed from \cite{2k_scenarios}. 

The loading for the 24,000-bus case as-provided is about 185 GW with the unrealistic assumption that all the generators are in-service. Hence at this initial state, no GSUs are associated with disconnected generators. To set a more realistic operating point for the provided high load scenario about 1500 generators are opened, providing a still high but more reasonable capacity reserve of about 30\%. A case with 70\% loading and nearly 3000 generators offline is also prepared for comparison.

Next, the GMD results for the grids are compared either assuming the GSUs for the decommitted generators are in-service (GSUs in) or out-of-service (GSUs out). Then for both conditions the grids are modeled using an assumed uniform electric field of 8 V/km in an eastward direction. To represent a particularly intense storm,  alpha and beta scaling factors are not used. Numerical results are presented first for the 2000-bus grid, then for the 24,000-bus grid with supplemental visualizations provided for additional detail.

%% file: Sections/IV_ResultsAnalysis.tex
\subsection{Reactive Power Losses}
For the high and low load cases, the total transformer reactive power losses decreased by 493 and 915 Mvar, respectively, when the GSUs were correctly modeled as out-of-service.  However, this presents an unfair comparison, since naturally one would expect that removing transformers from a network model should decrease the total reactive power consumption. Thus, the reactive power losses on only the transformers that are in-service in both scenarios are analyzed and presented in Table \ref{tab:Xfmr_Qloss_All}. Note that the reactive power losses on  GSUs for out-of-service generators are not included in the values in this table. Modeling GSU transformers for out-of-service generators as in-service yields errors of approximately 3\% and 16\% for the high and low load cases, respectively. 


Table \ref{tab:2K_Qloss_Area} shows a more granular representation of the reactive power losses for the 2000-bus grid by presenting the error in each of the eight system areas. For the high load case, the smallest discrepancy was in the West area, where calculated values with incorrect representation of GSU status yielded an error of 1.6\%. The largest error for the high load case was 7.3\% in the Far West. The low load case demonstrated a wider range, with a minimum error of 0.5\% in the West and a maximum error of 11.6\% in the South Central area.

\begin{table}[h!]
\captionsetup{justification=centering, labelsep=newline}
\caption{Transformer Reactive Power Losses on 2000-Bus Grid}
\label{tab:Xfmr_Qloss_All}
\begin{tabular}{l|rr}
\hline

                 & \multicolumn{1}{l}{High Load} & \multicolumn{1}{l}{Low Load} \\ \hline
\multicolumn{1}{c|}{GSUs in}   & 7557  & 3796   \\
\multicolumn{1}{c|}{GSUs out}  & 7806  & 4495   \\
\multicolumn{1}{c|}{Difference}   & 249   & 698    \\
\multicolumn{1}{c|}{\% Error}& 3.2\% & 15.5\% \\ \hline
\end{tabular}
\end{table}

\begin{table}[h!]
\captionsetup{justification=centering, labelsep=newline}
\caption{2000-Bus Reactive Power Loss \% Error by Area}
\label{tab:2K_Qloss_Area}
\begin{tabular}{ccc}
\hline
Area          & \multicolumn{1}{c}{\makecell{High Load Case \\Error [\%]}} & \multicolumn{1}{c}{\makecell{Low Load Case \\Error [\%]}} \\ \hline
Far West      & 7.3 & 4.4 \\
North         & 3.5 & 8.1            \\
West          & 1.6 & 0.5     \\
South         & 5.0 & 3.6    \\
North Central & 2.7 & 2.9              \\
South Central & 4.2 & 11.6             \\
Coast         & 2.0 & 4.9                 \\
East          & 3.0 & 9.3        \\ \hline
\end{tabular}
\end{table}




\subsection{Geomagnetically Induced Current Flows}

\subsubsection{Transformer GIC Effective}
 The effective GIC represents the combined effect of all the GICs flowing through the transformer windings. This value is used in the identification of transformers at risk for thermal damage. According to the NERC standard, transformers that have per phase effective GICs of over 75 A are at the greatest risk of thermal damage and, hence, should be reported by utilities to NERC \cite{nerc_Thermal}. Table \ref{tab:Xfmr_GIC_Eff} shows a comparison of the number of transformers with the effective GICs meeting this value between scenarios for the 2000-bus case. In both loading conditions, the number of GSUs with effective GIC flow greater than 75 A is higher in the scenario in which the GSUs are out-of-service for out-of-service generators. Thus, to correctly compute the effective transformer GIC flows requires accurate modeling of the GSU statuses, and can result in an increase in reportable findings to NERC.  
 
\begin{table}[h]
\captionsetup{justification=centering, labelsep=newline}
\caption{Number of Transformers with Per Phase\\ Effective GIC Above 75 A in 2000-Bus Grid}
\label{tab:Xfmr_GIC_Eff}
\begin{tabular}{c|cc}
\hline
\multicolumn{1}{l|}{} & \multicolumn{1}{l}{High Load} & \multicolumn{1}{l}{Low Load} \\ \hline
\multicolumn{1}{c|}{GSUs in}     & 46            & 44          \\
\multicolumn{1}{c|}{GSUs out} &  48            & 52          \\ \hline
\end{tabular}
\end{table}

\subsubsection{Transformer Neutral GICs}
When comparing the differences between the scenarios for both cases, both increases and decreases were observed in the GIC flows in the neutral connections for all transformers in the network. In the 2000-bus case, the average absolute change in GIC flow was around 7 A for the high load case with a standard deviation of 32 A, and 28 A for the low load case with a standard deviation of 64 A. The largest absolute change observed between scenarios for the 2000-bus high load case was 447 A. The largest absolute change observed between scenarios for the 2000-bus low load case was 432 A.

\subsubsection{Branch GICs}
\begin{figure}[t!]
    \centering
    \includegraphics[width=\textwidth]{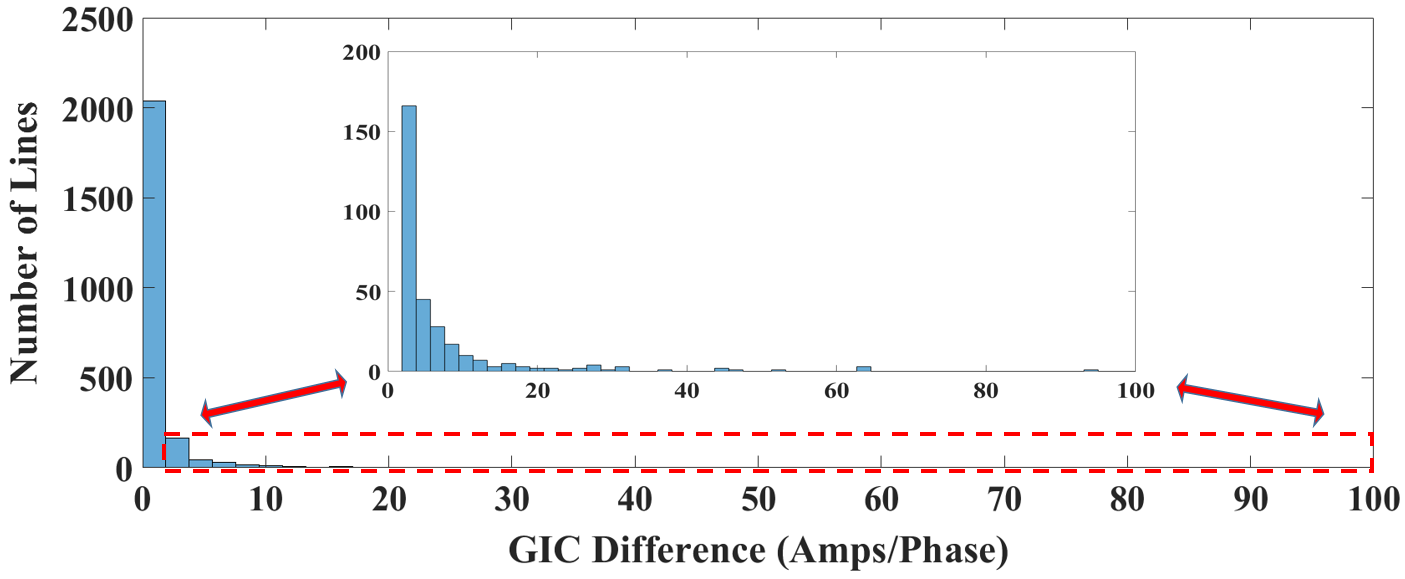}
    \caption{2000-bus low load branch GIC flow absolute error.}
    \label{Lowload_GIC_Branch}
\end{figure}

Incorrect modeling of the GSU transformer status also yielded errors in the calculated GIC flows on transmission lines. For the 2000-bus case in low load conditions, the distribution of the absolute differences between scenarios are shown for the low load case in Figure \ref{Lowload_GIC_Branch}. The average difference in calculated GIC flows on branches between the scenarios was a modest 0.5 A per phase and a majority of branches had errors of smaller magnitudes, however some branches displayed large errors in the GIC flows when incorrect GSU-status modeling was used. The most substantial difference of calculated GIC flows on a branch was 95 A per phase. 

\begin{figure}[t!]
    \centering
    \includegraphics[width=\textwidth]{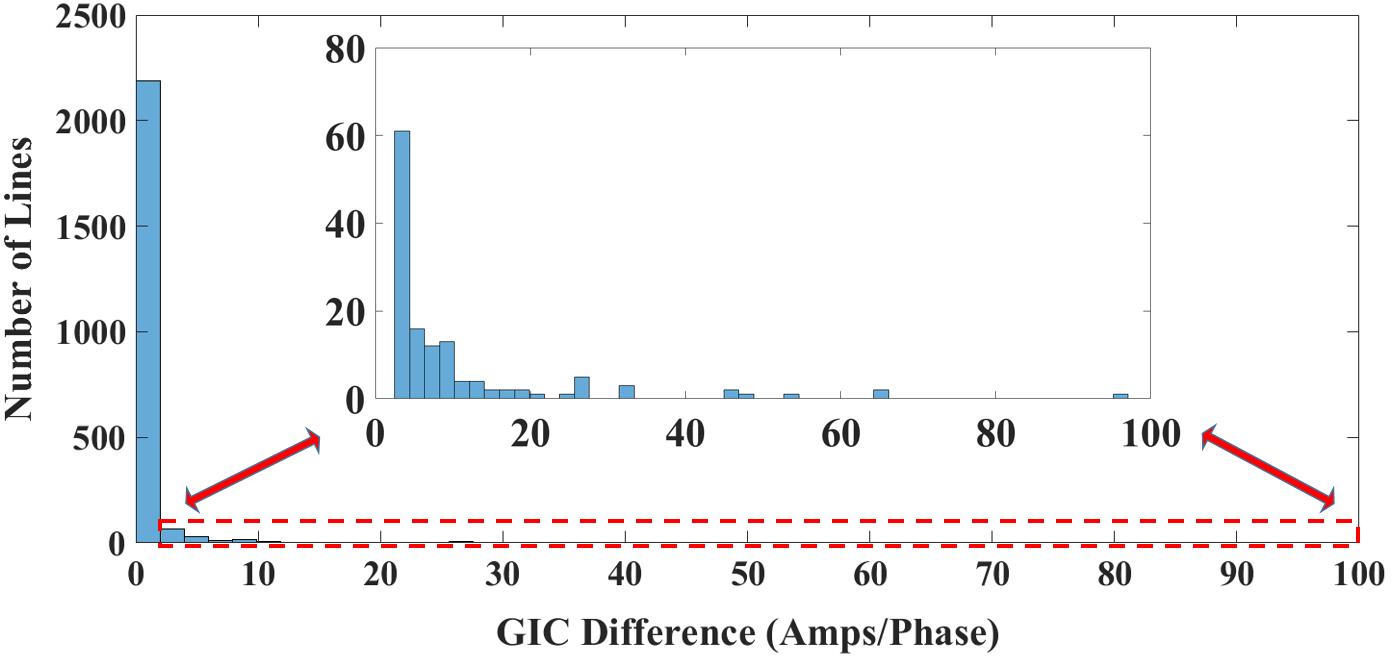}
    \caption{2000-bus high load branch GIC flow absolute error.}
    \label{Highload_GIC_Branch}
\end{figure}

The distribution of absolute error of branch GIC flow for the 2000-bus high load case is shown in Figure \ref{Highload_GIC_Branch}. The average error is approximately 0.2 A  per phase and the majority of the branch errors were within 0.1 A per phase of the average. The most substantial error observed was 96 A per phase. This shows that although the average change is fairly small, there are still substantial changes in GIC flows observed when the status of the GSUs are modeled correctly.

\subsection{Visualization of GIC Flows on 24,000-Bus Case}

\begin{figure} [h!]
    \includegraphics[width=\textwidth]{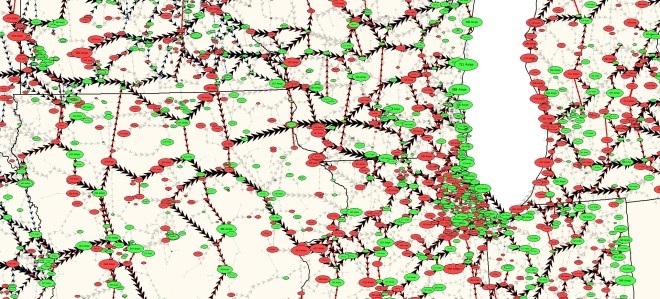}
    \caption{Visualization of GICs for 24,000-bus high load scenario with all GSUs in-service and an eastward field.}
    \label{O2}
\end{figure}

Figure \ref{O2} provides a visualization of the calculated GICs for a portion of the 24,000-bus high load scenario with an 8 V/km eastward field and all the GSUs unrealistically assumed to be in-service. In the figure, the GIC flows on the transformer neutral connections are shown using the geographic data view (GDV) visualization approach from \cite{O24k4} in which the size of the ovals is proportional to the GIC ground flow and the color indicates the direction (green into the ground and red out of the ground). The GICs in the transmission grid are visualized using black arrows superimposed on the lines with the size and direction of the arrow proportional to the flow \cite{O24k5}. While the GIC flow is certainly complicated, with overall flow direction from west to east is apparent, with grid boundaries (such as Lake Michigan) acting as barriers to the GIC flow. Similar visualizations could be created for the scenario with the GSUs out-of-service.             

\begin{figure}
    \includegraphics[width=\textwidth]{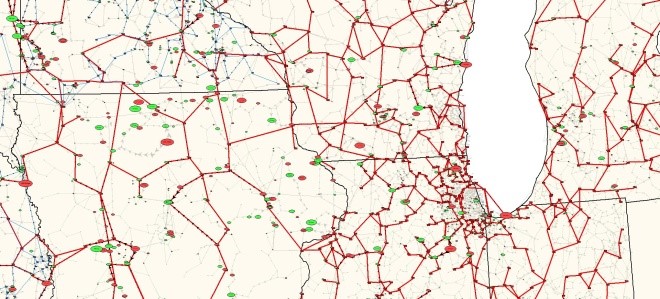}
    \caption{Visualization of GICs differences between GSUs in-service and out-of-service in 24,000-bus high load case.}
    \label{O3}
\end{figure}

In comparing the results between these different high load scenarios for the grid as a total the differences are relatively modest. For example there is about a 5\% difference in the total GIC losses for the entire case. However, the differences become more apparent as the focus shifts to say individual areas or individual components. At the area-level the largest difference is 12\% (in South Dakota) between the two scenarios. Figure \ref{O3} repeats the previous visualization, except now values shown are the differences between the GSU in-service and GSU out-of-service scenarios using the same scaling as Figure \ref{O2}. Note that the differences in the ground flows are fairly uniformly distributed across the region. 

\begin{figure}[h!]
    \includegraphics[width=0.9\textwidth]{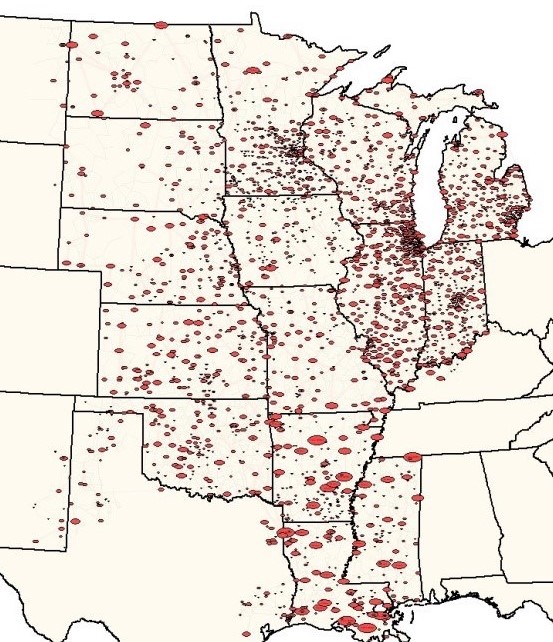}
    \caption{Visualization of GIC-induced substation reactive power losses assuming all GSUs in-service  in 24,000-bus high load case.}
    \label{O4}
\end{figure}

Naturally as the assumed load is decreased and the number of disconnected generators increases, the differences become more pronounced. The second part of this example compares the results when the load is reduced to 70\% of the previous case and the number of open generators is increased to almost 3000. An 8 V/km uniform electric field is again simulated (this time in a northward direction) for both the unrealistic but common power flow assumption of all the GSUs in-service, and for the more realistic case of the GSUs for the open generators out-of-service. Figure \ref{O4} visualizes the grid using the previous GDV method, except now each GDV oval shows the GIC-induced reactive power losses for the substations; the red color is used to indicate all the losses are positive.

 \begin{figure}[h!]
    \includegraphics[width=0.9\textwidth]{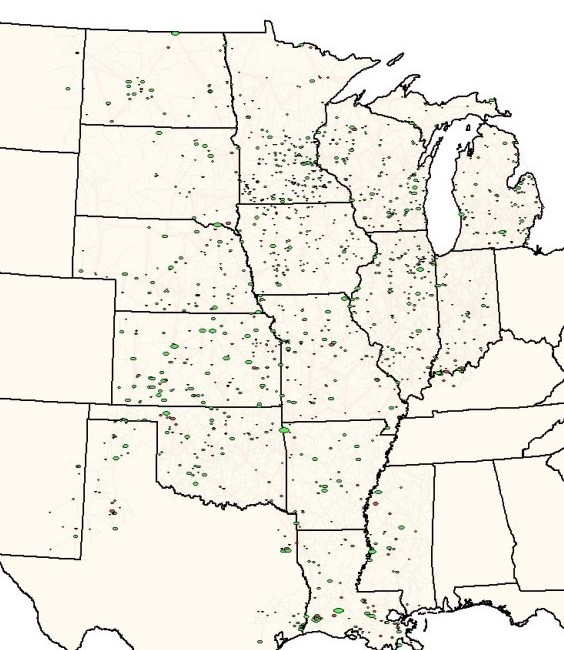}
    \caption{Visualization of GICs differences between GSUs in-service and out-of-service in 24,000-bus high load case.}
    \label{O5}
\end{figure}

Next, as before the scenario is repeated except with the GSUs associated with the out-of-service generators modeled as open. Now the total grid reactive power losses decrease by about 9\% with again the percentage varying more by area, substation and individual transformer. Figure \ref{O5} shows the variation between the two cases in the reactive power losses by substation using the same size scale as Figure \ref{O4}; a green color is used to show locations in which the losses decreased and red where they increased. 

Naturally with the GSUs opened the reactive power losses decreased in many transformers, of course going to zero for the thousands of now open GSUs. Overall the changed in the transformers with decreased losses represented 22\% of the original losses. However, this was partially offset by a net increase of 13\% for the transformers with increased losses. For a small number of transformers these changes are quite high, with increases of more than 100 A of effective GIC current for some. 

An important additional consideration of correctly modeling the statuses of the open generators GSUs is the GIC studies will become much more dependent on the generators that are assumed to be in-service. This means that two operating states with identical loads and assumed electric fields could have substantially difference GIC flows and associated reactive power losses.

Note that in presenting examples with either all the GSUs in-service or all out-of-service the implication is not that these would be the correct grid models. There are certainly situations in which the transformer connecting generation to the transmission grid might normally be in-service under almost all conditions, such as for a wind farm. The assumed all or nothing scenarios are just to demonstrate the importance of getting the GSU statuses correct when doing GIC related power flow studies.

%% file: Sections/V_SummaryFutureWork.tex
This paper has shown that GSU transformer status must be correctly modeled in order to correctly calculate the impacts of GICs. GSU transformer statuses are commonly overlooked as they have little to no impact on the steady state power flow results; however this paper shown that the GSU transformer statuses do impact the results of GIC studies. 
In this paper, the effects of considering the GSU transformer status in GMD studies are studied by comparing scenarios in which the GSU transformers for offline generators were either in-service or, more realistically, out-of-service. Simulation of GMD events on 2000- and 24,000-bus synthetic grids with various loading conditions showed that the modeling discrepancy resulted in calculation errors of up to  15.5\% for reactive power losses  and differences in GIC flows calculated on the transformers and branches as well. 

An important additional consideration of correctly modeling the statuses of the disconnected generators' GSUs is that GIC studies will become much more dependent on which set of generators are in-service. This means that a case with two different unit commitment scenarios but with identical load levels and assumed GMD electric fields could have substantially different GIC flows and associated reactive power losses. However, since both increases and decreases of GIC flow values were observed when generators were decommitted and their GSUs were disconnected, one cannot assume that a peak load case with most of the generators connected is the worst-case scenario. Thus, additional research needs to be done to determine the worst-case generator dispatch scenarios for calculating GIC flows in GMD simulations. 
